\documentclass[submission,copyright,creativecommons, shorttitle=Higher-Order TLA]{eptcs}
\providecommand{\event}{ACT 2019} 
\usepackage{breakurl}             
\usepackage{underscore}           

\pdfoutput=1
\usepackage[utf8]{inputenc}
\usepackage[english]{babel}
\usepackage[T1]{fontenc}
\usepackage{graphicx}				
\usepackage{bbold}
\usepackage{amssymb}
\usepackage{stmaryrd}

\usepackage{tikz,tikz-cd}
\usepackage{mathpartir}
\usepackage{amsmath}
\usepackage{ latexsym }
\usepackage{cmll}
\usepackage{amsthm}
\usepackage{listings}
\usepackage{kotex}
\usepackage{hyperref}
\usepackage{bm}
\usepackage{tikz}
\usepackage{lipsum}

\newtheorem{theorem}{Theorem}

\newtheorem{proposition}{Proposition}
\newtheorem{definition}{Definition}

\newtheorem{lemma}{Lemma}

\newcommand{\msubst}[1]{[{#1}]}
\newcommand{\trans}[1]{\llbracket{#1}\rrbracket}

\newcommand{\alwaysM}[1]{\Box {#1}}

\newcommand{\alwaysAction}[2]{\Box [{#1}]_{{#2}}}
\newcommand{\forallFlex}{\bm{\forall}\kern-0.6em\bm{\forall}}
\newcommand{\existsFlex}{\bm{\exists}\kern-0.6em\bm{\exists}}

\begin{document}
\title{Topos Semantics for a Higher-order Temporal Logic of Actions}
\date{} \author{Philip Johnson-Freyd
\institute{Sandia National Laboratories}
\email{pajohn@sandia.gov}
\and
Jon Aytac
\institute{Sandia National Laboratories}
\email{jmaytac@sandia.gov}
\and
Geoffrey Hulette
\institute{Sandia National Laboratories}
\email{ghulett@sandia.gov}}
\def\titlerunning{Topos Semantics for a Higher-order Temporal Logic of Actions}
\def\authorrunning{Philip Johnson-Freyd, Jon Aytac, Geoffrey Hulette}

\maketitle

\begin{abstract}
TLA is a popular temporal logic for writing stuttering-invariant
specifications of digital systems.  However, TLA lacks higher-order
features useful for specifying modern software written in higher-order
programming languages.  We use categorical techniques to recast a
real-time semantics for TLA in terms of the actions of a group of time
dilations, or ``stutters,'' and an extension by a monoid incorporating
delays, or ``falters.''  Via the geometric morphism of the associated
presheaf topoi induced by the inclusion of stutters into falters, we
construct the first model of a higher-order TLA.
\footnote{
Sandia National Laboratories is a multimission laboratory managed and
operated by National Technology \& Engineering Solutions of Sandia,
LLC, a wholly owned subsidiary of Honeywell International Inc., for
the U.S. Department of Energy's National Nuclear Security
Administration (NNSA) under contract DE-NA0003525. This work was
funded by NNSA's Advanced Simulation and Computing (ASC) Program. This
paper describes objective technical results and analysis. Any
subjective views or opinions that might be expressed in the paper do
not necessarily represent the views of the U.S. Department of Energy
or the United States Government.
}
\end{abstract}

\section{Introduction}
\label{sec:introduction}

The Temporal Logic of Actions (TLA) is a temporal logic commonly
used for specifying digital computer
systems~\cite{Lamport:1994:TLA:177492.177726, newcombe-amazon}. TLA
formulae are linear temporal properties \emph{invariant under
``stuttering.''} Stuttering invariant specifications written
as TLA formulae are easily composed, using
nothing more than conjunction, with no implicit assumptions about
synchronization.  Stuttering invariance also leads to a simple but
powerful notion of ``refinement,'' that is, showing that a detailed
specification implements an abstract one.

In~\cite{Lamport:1994:TLA:177492.177726} Lamport presents TLA as a
first-order logic, but, in specifications, higher-order features are
often desirable.  For example, one would often like to prove a rule of
inference that works over \emph{all} propositions or \emph{all}
predicates.  Lamport must introduce special syntax (e.g., for
fairness) where in a higher-order context these language features
could be replaced with simple functions on propositions.
Moreover, programmers today often work in higher-order
programming languages and the powerful abstraction features in these
languages (e.g., a generalized ``map'' function) are not easily
expressed in TLA specifications.

As a step towards the goal of defining a higher-order TLA, we
present \emph{a} model in which it could be interpreted.  In standard
linear temporal logics, which do not feature stuttering invariance,
higher-order features can be modeled in the so-called “Topos of Trees”
(i.e., presheaves over
$\omega$)~\cite{Jeltsch:2012:TCC:2379837.2379873,riba:hal-01620992}.
Another impressive line of work on ``Temporal Types'' takes a topos theoretic
approach based on translation invariant sheaves (using the additive structure of
$\mathbb{R}$) \cite{schultzSpivakTemporalTypes}.
Unfortunately, these models cannot capture TLA's stuttering invariance.


Our categorical model of higher-order TLA meets several desiderata,
motivated by the observations above:
\begin{enumerate}
\item it should provide a model of higher-order classical S4 (TLA is a special
  case of this modal logic);
\item it should have a ``temporal'' interpretation which accounts for stuttering
  invariance;
\item it should correspond with an equivalent notion of validity, in the first-order subset, to the standard semantics of TLA.
\end{enumerate}

We believe our model to be the first that is suitable for a
higher-order TLA. It is constructed as follows. First, we switch
perspective, from the standard discrete-time semantics of TLA to an
alternative real-time semantics found in the
literature~\cite{DBLP:journals/logcom/KaminskiY03} and reviewed in
Section~\ref{sec:tla-background}.  Then, recalling that models for
higher-order modal logic can be generated by geometric morphisms
between topoi (Section \ref{sec:hol-background}), we construct our
model by recasting the real-valued semantics by way of such a
geometric morphism (Section~\ref{sec:the-model}). Our key insight was
to consider stutterings as a group,
leading to a generalization of stuttering, which we call
``faltering.''

\section{The Temporal Logic of Action}
\label{sec:tla-background}

Like Pnueli's Linear Temporal Logic
(LTL)~\cite{Pnueli:1977:TLP:1382431.1382534}, TLA adopts the
perspective of linear time: formulae classify sets of (linear)
infinite traces of a system evolving through time. Also like LTL, TLA
has temporal modalities ``always'' ($\Box$) and ``eventually''
($\Diamond$). However, unlike LTL, TLA has no ``next'' ($\circ$)
modality. Instead, TLA has a notion of ``actions'' that describe
instantaneous changes in the system state, but which also allow
``stuttering steps'' in which the trace evolves in time but the state
remains unchanged.  Thus, unlike LTL, TLA formulae are always
``stuttering invariant,'' that is, they cannot differentiate traces by
how long they stutter.


\begin{figure*}[t]
  \begin{gather*}
  \begin{aligned}
    E \in \textrm{Terms} &::= x \mid \mathbf{x}
      \mid \mathbf{x'}
      \mid f(E_1,\ldots,E_n) \\
    A \in \textrm{Actions} &::= R(E_1,\ldots,E_n)
      \mid E_1 = E_2
      \mid \forall x.A
      \mid A_1 \land A_2
      \mid \lnot A \\
    P \in \textrm{Propositions} &::= \alwaysM{P}
      \mid \lnot P
      \mid P_1 \land P_2
      \mid \alwaysAction{A}{< \mathbf{x_1},\ldots,\mathbf{x_n} >}
      \mid \forall x.P
      \mid \forallFlex \mathbf{x}.P \\
    T \in \textrm{Formulae} &::= A \mid P
  \end{aligned}~~
  \begin{aligned}
    \exists x.T &\triangleq \lnot \forall x.\lnot T  \\
    \existsFlex \mathbf{x}.P &\triangleq
                \lnot \forallFlex \mathbf{x}.\lnot P \\
    T_1 \lor T_2 &\triangleq \lnot (\lnot T_1 \land \lnot T_2) \\
    T_1 \Rightarrow T_2 &\triangleq \lnot T_1 \lor T_2
  \end{aligned}
  \end{gather*}
  \caption{Syntax and Syntactic Sugar of TLA}
  \label{fig:tla-syntax}
\end{figure*}

Syntactically, TLA has two classes of formulae
(Figure \ref{fig:tla-syntax}): actions, which denote instantaneous
changes to the system state, and temporal formulae, which are
predicates on traces.

Actions are normal first-order logic formulae except in the handling
of terms. Variables appearing in terms can be ``rigid'' (written in
italics), indicating that they do not change over time, or
``flexible'' (written in bold face), indicating that they
may. Flexible variables may appear primed ($\mathbf{x'}$) or unprimed
($\mathbf{x}$) denoting the variable's value in the next or current
state, respectively.

Temporal formulae are comprised of the usual propositional connectives
and temporal quantifiers, along with a special operator
$\alwaysAction{A}{v}$, where $A$ is an action and $v$ is a function
on the system state). Intuitively, the formula $\alwaysAction{A}{v}$
means ``it is always the case that either the action $A$ happens or
$v$ does not change.'' TLA is also equipped with ordinary
(first-order) quantifiers over rigid variables $\forall x.P$ as well
as ``temporal'' quantifiers over flexible variables
$\forallFlex \mathbf{x}.P$.

\begin{figure*}[t]
\begin{gather*}
\begin{aligned}
\trans{x}(\theta,\sigma,\sigma') &= \theta(x) \\
\trans{\mathbf{x}}(\theta,\sigma,\sigma') &= \sigma(x) \\
\trans{\mathbf{x'}}(\theta,\sigma,\sigma') &= \sigma'(x) \\
\end{aligned}\\
\begin{aligned}
\trans{f(E_1,\ldots,E_n)}(\theta,\sigma,\sigma') &=
\mathcal{F}(f)(\trans{E_1}(\theta,\sigma,\sigma'),\ldots,\trans{E_1}(\theta,\sigma,\sigma'))
\end{aligned}\\
\begin{aligned}
\theta,\sigma,\sigma' &\models A_1 \land A_2
&\textrm{iff}&~
        (\theta,\sigma,\sigma' \models A_1)~
        \textrm{and}~
        (\theta,\sigma,\sigma' \models A_1)\\
\theta,\sigma,\sigma' &\models \lnot A
&\textrm{iff}& ~
\theta,\sigma,\sigma' \not \models A\\
\theta,\sigma,\sigma' &\models E_1 = E_2
&\textrm{iff}&~
\trans{E_1}(\theta,\sigma,\sigma') = \trans{E_2}(\theta,\sigma,\sigma')\\
\theta,\sigma,\sigma' &\models \forall x.A
&\textrm{iff}&~
\textrm{for every $v \in \mathcal{D}$ } (\theta \uplus x \mapsto v), \sigma, \sigma' \models A\\
\theta,\sigma,\sigma' &\models R(E_1,\ldots,E_n)
&\textrm{iff}&~
\mathcal{R}(R)(\trans{E_1}(\theta,\sigma,\sigma'),\ldots,\trans{E_n}(\theta,\sigma,\sigma'))
\end{aligned}~~\\
\begin{aligned}
\theta,\rho &\models P_1 \land P_2
&\textrm{iff}&~
\theta,\rho \models P_1~\textrm{and}~\theta,\rho \models P_2\\
\theta,\rho &\models \lnot P
&\textrm{iff}&~
\theta,\rho \not \models P\\
\theta,\rho &\models \alwaysM{P}
&\textrm{iff}&~
\textrm{for every $n \in \mathbb{N}$ } \theta,\rho[n,\ldots] \models P \\
\theta,\rho &\models \alwaysAction{A}{< \mathbf{x_1},\ldots,\mathbf{x_m} >}
&\textrm{iff}&~
\textrm{for each $n \in \mathbb{N}$ either } \theta,\rho[n],\rho[n+1] \models A\\
&&&\textrm{ or } \forall i \in [1,m]. \rho[n](\mathbf{x_i})=\rho[n+1](\mathbf{x}_{i})\\
\theta, \rho &\models \forall x.P
&\textrm{iff}&~
\textrm{for every $v \in \mathcal{D}$ } (\theta \uplus x \mapsto v), \rho \models P\\
\theta, \rho &\models \forallFlex \mathbf{x}.P
&\textrm{iff}&~
\textrm{for every $d \in \mathcal{D}^\mathbf{N}$ and $\rho' \approx \rho$}, \theta, \rho' \uplus (\mathbf{x} \mapsto d) \models P
\end{aligned}
\end{gather*}
\caption{Discrete Time Semantics of TLA}
\label{fig:tla-lamport-semantics}
\end{figure*}

Lamport's semantics for TLA (Figure \ref{fig:tla-lamport-semantics})
interprets temporal formulae using a discrete model of time.
Traces are modeled as functions from
natural numbers to a ``state,'' where states are assignments of values
for each flexible variable.


Lamport's semantics are unusual in the handling of the flexible
quantifier ($\forallFlex$). Na\"ively, flexible quantification would be
\begin{align*}
  \theta, \rho \models \forallFlex \mathbf{x}.P ~~~ \textrm{iff}~~ 
  \textrm{for every $d \in \mathcal{D}^\mathbb{N}$}, \\ \theta, \rho \uplus (\mathbf{x} \mapsto d) \models P
  \end{align*}

  \begin{definition}[Discrete Stuttering Equivalence]
  Given any set $S$, two behaviors $\rho_{1}$, $\rho_{2}$ are said to be stuttering
  equivalent if there exists monotone surjections
  $\phi_{1},\phi_{2}:\mathbb{N}\to\mathbb{N}$ such that $\phi_{1}^{\star}\rho_{1}=\phi_{2}^{\star}\rho_{2}$    
\end{definition}

Unfortunately, in this semantics, the definition of flexible
quantification must explicitly ``bake in'' stuttering invariance (see the
semantics of $\forallFlex$ in Figure \ref{fig:tla-lamport-semantics}) and
this makes flexible quantification behave quite differently from the
ordinary semantics of modal logic.

\begin{proposition}[Stuttering Equivalence of TLA]
For any $P,\theta,\rho,\rho'$ such that $\rho \approx \rho'$
\[ \theta,\rho \models P \textrm{ iff } \theta,\rho' \models P. \]
\end{proposition}

Kaminski and Yariv \cite{DBLP:journals/logcom/KaminskiY03} proposed an
alternative semantics for TLA based on a continuous notion of time. In
this setting traces are interpreted as ``non-Zeno'' functions from the
non-negative real numbers.

\begin{definition}[Non-Zeno function]
A non-Zeno function over a set $S$ is a function $f$ from
non-negative real numbers to $S$ such that
\begin{enumerate}
\item for every $t \in \mathbb{R}_{\geq 0}$ there exists a positive
$\epsilon$ such that for all $t'$ where $t \leq t' < t
+ \epsilon$ we have $f(t)=f(t')$ and
\item there is no bounded increasing sequence $t_0,t_1,t_2,\ldots$ such that
forall $i$, $f(t_i) \neq f(t_{i+1})$.
\end{enumerate}
\end{definition}

These two conditions ensure that a non-Zeno function does not change
too quickly: the first condition guarantees that each state is held
for positive time, while the second ensures that only a finite number
of states are visited in any finite length of time. We (ab)use the
notation $S^{\mathbb{R}^{+}}$ to refer to the set of non-Zeno
functions over $S$.

Stuttering invariance of a set of such non-Zeno functions is modeled
as closure under pre-composition by homeomorphisms on
$\mathbb{R}_{\geq 0}$ (with the standard topology). The alternative
continuous semantics (Figure \ref{fig:tla-continuous-semantics})
yields exactly the same notion of truth as Lamport’s original
semantics, while avoiding the need to ``bake in'' stuttering
invariance in its definitions.

\begin{figure*}
\begin{gather*}
\begin{aligned}
next(\tau,S) & \triangleq 0 
&\textrm{ when } \forall t \in \mathbb{R}_{\geq 0}, \forall x \in S, \tau(0)(x) = \tau(t)(x) \\
next(\tau,S) & \triangleq \textbf{sup}\{r \mid \forall 0 \leq k \leq r, \forall x \in S, \tau(0)(x) = \tau(k)(x)\}
&\textrm{ otherwise}
\end{aligned}\\
\begin{aligned}
\theta, \tau &\models_{\mathbb{R}} \alwaysAction{A}{\mathbf{x}_1,\ldots,\mathbf{x}_n}
&\textrm{iff}&~
r = 0
\textrm{ or }
\theta, \tau(0), \tau(r) \models A \\
&&& \textrm{ where } r = next(\tau,\{\mathbf{x}_i | 0 \leq i \leq n\})\\
\theta, \tau &\models_{\mathbb{R}} T_1 \land T_2
&\textrm{iff}&~
\theta, \tau \models_{\mathbb{R}} T_1 ~\textrm{and}~ \theta, \tau \models_{\mathbb{R}} T_2\\
\theta, \tau &\models_{\mathbb{R}} \lnot T
&\textrm{iff}&~
\theta, \tau \not \models_{\mathbb{R}} T\\
\theta, \tau &\models_{\mathbb{R}} \forall x . T
&\textrm{iff}&~
\textrm{for every}~ v \in \mathcal{D} ~\textrm{we have}~
(\theta, x \mapsto v), \tau \models_{\mathbb{R}} T\\
\theta, \tau &\models_{\mathbb{R}} \forallFlex \mathbf{x} . T
&\textrm{iff}&~
\textrm{for every}~ v \in \mathcal{D}^{\mathbb{R}^+} ~\textrm{we have}~
\theta, (x \mapsto (\tau(r), \mathbf{x} \mapsto v(r))) \models_{\mathbb{R}} T\\
\theta, \tau &\models_{\mathbb{R}} \alwaysM{T}
&\textrm{iff}&~
\textrm{for every}~ k \in \mathbb{R}_{\geq 0} ~\textrm{such that}~ \theta, \tau[k..] \models_{\mathbb{R}} T\\
\end{aligned}
\end{gather*}
\caption{Continuous-time Semantics of TLA}
\label{fig:tla-continuous-semantics}
\end{figure*}

This continuous semantics clarifies many aspects of TLA. It explains
stuttering invariance as invariance under time dilation. Furthermore,
it presents rigid and flexible variables uniformly, allowing them to
be viewed as coming from two different types. Categorically, this
means rigid and flexible quantification should correspond to
quantification over different objects.
\section{Semantics of Higher-order Logic}
\label{sec:hol-background}

Higher-order Logic (HOL) (see \cite{church1940formulation}) combines a
(possibly intuitionistic) logic with the simply-typed
$\lambda$-calculus. It may be viewed as an extension to multi-sorted
first-order logic that adds features for quantifying over function
types and propositions.

Modal variants of higher-order logic are usually formed simply by
adding additional modal operators exactly as one would in a
propositional logic.

There are many semantics for higher-order logic. In the ``standard''
semantics, types are interpreted as sets, function types are
interpreted as the set of all functions between their constituents,
and propositions are interpreted as booleans. This model is
incomplete, however.

A more general class of model is found in topoi.

\begin{definition} A topos is a cartesion closed category $\mathcal{E}$
  possesing all finite limits and a subobject classifier,
  i.e. an object $\Omega$ and
      a monic arrow $\text{\sf true}:1\to\Omega$ such that $\forall$ monic
      $m:S\to B$  $\exists! \phi_{m}:B\to\Omega$ 
      such that
      \begin{tikzcd}
      S\arrow[tail]{d}{}{m}\arrow{r}{!}&1\arrow{d}{\text{\sf true}}\\
      B\arrow[dashed]{r}{}{\phi_{m}} & \Omega
    \end{tikzcd} is a pullback.
\end{definition}

In the na\"ive topos semantics, types (also, contexts) are interpreted
as objects, terms are interpreted as morphisms, function types are
interpreted by way of the inner hom, and the proposition type is
interpreted as the subobject classifier.

However, this topos semantics is still too strong---it justifies
additional laws which are not derivable from the natural deduction
rules in Figure~\ref{fig:iHOL}. In particular, the topos semantics
imposes upon higher-order logic the additional property of
extensionality of entailment (see~\cite{jacobs1999categorical} 5.3.7)

\begin{mathpar}
\inferrule{\Gamma\vdash
P,Q:\sigma\to\textrm{Prop} \\
\Gamma,x:\sigma\mid\Theta,Px\vdash Qx \\
\Gamma,x:\sigma\mid\Theta, Qx\vdash Px} {\Gamma\mid\Theta\vdash P=_{\sigma\to\textrm{Prop}}Q}
\end{mathpar}.

A class of categorical models for higher-order logic with more
examples is obtained by weakening the structure involved in the
subobject classifier.

\begin{definition}[Hyperdoctrine] Let
  $\text{\sf P}:\textbf{\sf C}^{op}\to\textbf{\sf HeyAlg}$
   be a functor from a cartesion closed $\textbf{\sf C}$
   into the category of Heyting algbras
  such that:
  \begin{enumerate}
  \item $\forall X,Y:\textrm{Obj}\textbf{\sf C}$
    there are monotone
  $\exists^{X}_{Y},\forall^{X}_{Y}:\textrm{Hom}_{\textbf{\sf PreOrd}}(\text{\sf  P}(X\times Y),\text{\sf P}(Y))$
  such that for $\pi : X \times Y \to Y$ the projection
  $ \exists^{X}_{Y}\dashv\text{\sf P}(\pi)\dashv\forall^{X}_{Y} $
  and
  satisfying the Beck-Chevalley condition\\ $\forall f:Y\to Y'$ 
  \begin{tikzcd}
    \text{\sf P}(X\times Y') \arrow{d}{}[swap]{\forall^{X}_{Y'}}    \arrow{r}{\text{\sf P}(\textrm{id}_{X}\times f)}&\text{\sf P}(X\times    Y)\arrow{d}{}[swap]{\forall^{X}_{Y}}\\
   \text{\sf P}Y'\arrow{r}{}[swap]{\text{\sf P}f} & \text{\sf P}Y
  \end{tikzcd}
  commutes 
  as does the similar $\exists^{X}_{Y}$ diagram;
\item $(\textsf{Forget} \cdot \textsf{P}) : \textbf{\sf C}^{op} \to \textbf{\sf Set}$ is representable.
 \end{enumerate}
 \end{definition}

Hyperdoctrines provide a setting for a sound and complete semantics
for HOL by modeling contexts using the underlying cartesian closed
category structure, with the Heyting algebra of propositions over
those contexts given by the functor, and the quantifiers induced by
the adjoints.\footnote{Completeness, as is often the case,
holds for the \emph{class} of models by constructing an appropriate
\emph{syntactic} object initial in the category of hyperdoctrines as in
\cite{Lambek2011reflections}. }  Moreover, by replacing the category of Heyting algebras
with the category of Boolean algebras, we gain a notion of ``classical
hyperdoctrine,'' which provides a sound and complete semantics for
classical higher-order logic. Finally, using an even stronger category of
``modal algebras'' yields a model of S4 modal higher-order logic.

\begin{definition}
  A \textbf{modal algebra} is a pair $(A,\Box):\text{Obj}(\textbf{\sf MAlg})$
  where $A$ is a Heyting algebra and $\Box$ is a left exact comonad on $A$.

  A \textbf{modal algebra morphism} $f : (A,\Box) \to (B,\Box')$ is
  a morphism of the underlying Heyting algebras which commutes
  with the modalities in the sense that $f \cdot \Box = \Box' \cdot f$.
\end{definition}

 \begin{definition}[Modal Hyperdoctrine] Let
  $\text{\sf P}:\textbf{\sf C}^{op}\to\textbf{\sf MAlg}$ be a functor
   from a small cartesian closed category $\textbf{\sf C}$
   into the category of Modal algbras
   $\textbf{\sf MAlg}$ otherwise satisfying the axioms of a
   hyperdoctrine.
\end{definition}

The hyperdoctrine semantics fully generalizes the topos semantics, as
every topos $T$ induces a (intuitionistic) hyperdoctrine
\begin{equation}\label{eq:omega-hyperdoctrine}
(T,\textrm{Hom}_T(-,\Omega)).
\end{equation}

However, these are not the only hyperdoctrines of interest. Specifically,
the only fact about $\Omega$ in
equation \ref{eq:omega-hyperdoctrine} required
for the resulting structure to be a
(intuitionistic) hyperdoctrine is that it forms an internal complete
Heyting algebra in $T$.

Given any topos $\mathcal{E}$ and internal complete Heyting algebra
$H$ in $\mathcal{E}$, there is a natural way of equipping
$\textrm{Hom}_{\mathcal{E}}(-,H)$ with a Heyting algebra structure so
that $(\mathcal{E},\textrm{Hom}_{\mathcal{E}}(-,H))$ forms a
hyperdoctrine.


If $H$ is an internal complete boolean or modal algebra in $T$, then
the resulting hyperdoctrine will be classical or modal,
respectively~\cite{2014arXiv1403.0020A}.


In this topos-theoretic setting, we can apply a simple recipe for
constructing a topos together with internal complete modal
algebras. Recall

\begin{definition}
Let $\mathcal{E}$, $\mathcal{F}$ be topoi. A \textbf{geometric
    morphism} $\text{\sf f}:\mathcal{E}\to\mathcal{F}$ is an
    adjunction \begin{tikzcd} \mathcal{E}\arrow[shift
    left=1.5ex]{r}{\text{\sf f}_{\star}}[swap]{\top}
    & \mathcal{F}\arrow[shift left=1.5ex]{l}{\text{\sf
    f}^{\star}} \end{tikzcd} such that the left adjoint $\text{\sf
    f}^{\star}$, known as the inverse image, preserves finite
    limits. If every object $X:\textrm{Obj}(\mathcal{E})$ is a
    subquotient of an object of the inverse image $\text{\sf
    f}^{\star}$, so that there exists $Y:\textrm{Obj}(\mathcal{F})$
    and diagram $f^{\star}(Y)\leftarrowtail S\twoheadrightarrow X$,
    then $\textbf{\sf f}$ is \textbf{localic}.
\end{definition}

Geometric morphisms are a source of internal complete Heyting algebras.

\begin{proposition} Let $\text{\sf f}:\mathcal{E}\to\mathcal{F}$ a geometric
  morphism. Then $\text{\sf f}_{\star}(\Omega_{\mathcal{E}})$ is a complete
  Heyting algebra
  internal to $\mathcal{F}$.
\end{proposition}

Geometric morphisms are also a source of adjoint pairs of maps of
complete Heyting algebras.

\begin{lemma} [\cite{johnstone2002sketches} C1.3]
In any topos $\mathcal{E}$, the subobject classifier
$\Omega_{\mathcal{E}}$ is the initial complete Heyting algebra
object. That is, for all complete Heyting algebras $H$ internal to
$\mathcal{E}$, there is a unique map of complete Heyting algebras
$i:\Omega_{\mathcal{E}}\to H$. Moreover, the right adjoint of $\tau$
is the classifying map of the top element $\top_{H}:1\to H$.
\end{lemma}

This adjoint pair of maps defines a useful comonad.

\begin{lemma}[\cite{2014arXiv1403.0020A}]
  Given a complete Heyting algebra $H$ internal to topos $\mathcal{E}$,
  let $i\vdash\tau$ the canonical adjunction
  $i:\Omega_{\mathcal{E}}\stackrel{\rightarrow}{\leftarrow}H:\tau$.
  The composite $i\circ\tau$ is an $\textrm{S4}$ modality on $H$.
\end{lemma}

If we have \emph{two} topoi, $\mathcal{E}$ and $\mathcal{F}$, and a
geometric morphism $\text{\sf f} : \mathcal{E} \to \mathcal{F}$ then the image of
the subobject classifier of $\mathcal{E}$ in $\mathcal{F}$ is an
internal complete modal algebra in $\mathcal{F}$.


An illustrative example is given by a topos-theoretic view of Kripke
semantics. Let $K$ be a preorder, interpreted as a collection of
``possible worlds,'' together with an accessibility relation. By $|K|$
we mean the discrete category with the same underlying objects as $K$.

The inclusion $|K| \to K$ induces a geometric morphism $\text{\sf f} :
Psh(|K|) \to Psh(K)$.

\begin{lemma}[\cite{johnstone1981factorization}, prop. 3.1]
Let $\text{\sf f}:\text{\sf D}\to\text{\sf C}$ be a functor of small
categories.  If $\text{\sf f}$ is faithful, then the induced geometric
morphism $Psh(\textbf{\sf D})\to Psh(\textbf{\sf C})$ is localic.
\end{lemma}

Thus we obtain a modal hyperdoctrine on
$(Psh(K),\textrm{Hom}_{Psh(K)}(-,\text{\sf f}_{\star}(\Omega_{Psh(|K|)}))$.  In
particular, as $|K|$ is a groupoid, $\mathcal{E}=Psh(|K|)$ is a Boolean topos, so
$\text{\sf f}_{\star}(\Omega_{\mathcal{E}})$ is not only a complete
Heyting algebra internal to $\mathcal{F}=Psh(K)$, it is an
internal Boolean algebra!  The resulting logic is classical, even
though $Psh(K)$ is very much not a boolean topos in general (it is,
instead, a Kripke model of an intuitionistic logic). The
internal logic of this modal hyperdoctrine is, in the first-order
fragment, exactly what we would get from the Kripke semantics over $K$.
And thus we have a simple presentation of a higher-order version of
that semantics.
\section{The Model}
\label{sec:the-model}
Now we are ready to construct a candidate model for a Higher-order
TLA.

Why not simply use the topos-theoretic Kripke semantics, described in
Section~\ref{sec:hol-background}, applied to the discrete semantics?
This approach will fail because TLA’s discrete semantics is \emph{not}
an ordinary Kripke semantics, since flexible quantification is not
ordinary Kripke quantification (see
Section~\ref{sec:tla-background}). Even the continuous semantics is
not adequately captured in the ordinary, preorder-based, Kripke view
since Kripke does not account for stuttering.

We must build a model that includes stuttering invariance from the get
go. Pre-orders are inadequate to this task. Luckily, the geometric morphism
construction described in Section~\ref{sec:hol-background} is not
specific to Kripke's inclusion of a discrete set into a
preorder. \emph{Any} faithful functor between small categories whose
domain is a groupoid induces a model of classical higher-order modal
logic.

Our model is enabled by the following elementry observation: in the
continuous semantics, stuttering invariance is precisely closure under
the action the group of stutters.

\begin{definition}[Stutter]
  A stutter is a continuous function
  $\mathbb{R}_{\geq 0} \to \mathbb{R}_{\geq 0}$
  with continuous inverse.

  By $\mathcal{S}$ we denote the group of stutters
  \[ \mathcal{S} = (
  \{f : \mathbb{R}_{\geq 0} \to \mathbb{R}_{\geq 0} \mid
  f \textrm{ is a stutter}
  \}
  ,\cdot,
  id_{\mathbb{R}_{\geq 0}}
  ) \].
\end{definition}


We will adopt the convention of viewing any monoid $G$ as the category
$\text{\sf B}G$ with one object and one monoid's worth of
morphisms. This way the category of $G$-sets and $G$-set
morphisms for a group $G$ (more generally, for any monoid) is just
$Psh(\text{\sf B}G)$.

Non-Zeno functions over a set form a $\mathcal{S}$-set where the action
of $\mathcal{S}$ is pre-composition.  Stuttering invariant subsets of
that set are then, exactly, sub $\mathcal{S}$-sets. As such, the
category of $\mathcal{S}$-sets ($Psh(\text{\sf B}\mathcal{S}))$ seems
to be closely connected to our problem.
Since $\mathcal{S}$ is a group,
$\text{\sf B}\mathcal{S}$ is a groupoid, and the presheaf topos
$Psh(\text{\sf B}\mathcal{S})$ is boolean. Therefore, it is a tempting
target for the semantics of a higher-order TLA. We already know this will not
work on its own though, as a topos is not enough to interpret the
modalities. The most important modality for our purposes is $\Box$. A behavior
(viewed as a non-Zeno function) is always a member of some set of
behaviors if, given any initial delay in which the behavior is not
observed, the remainder is in that set. Thus, while stuttering
invariance has to do with closure under dilation of time by
bi-continuous functions, $\Box$ has to do with the translation of
time.

To that end, we introduce a generalization of stutters, which we call
``falters,'' which can include translation as well as dilation.

\begin{definition}
  A falter is a monotone function $f : \mathbb{R}_{\geq
  0} \to \mathbb{R}_{\geq 0}$ such that the function $ x \mapsto f(x)
  - f(0) $ is a stutter.

  By $\mathcal{F}$ we denote the monoid of falters
  (under function composition).
\end{definition}

There is a natural morphism of monoids
$\iota: \mathcal{S} \to \mathcal{F}$ given by inclusion, inducing a
faithful functor $\iota:\text{\sf B}S\to\text{\sf B}\mathcal{F}$. As
mentioned in Section~\ref{sec:hol-background}, such a faithful functor
induces a localic geometric morphism on the associated presheaf
categories $\iota^{\star}\dashv\iota_{\star}:Psh(\text{\sf
B}\mathcal{S})\stackrel{\to}{\leftarrow}Psh(\text{\sf
B}\mathcal{F})$.  Our proposed model for a higher-order TLA is the
hyperdoctrine induced by this geometric morphism.

We will now elaborate some details of this model. We consider
$\mathcal{F}$-sets to be ``temporal types'' as these are the types
about which we can talk in our model. The type of flexible variables
over some base set are computed according to the functor
\begin{align*}
\textrm{Flex} &: \textrm{Set} \to Psh(\text{\sf B}\mathcal{F})\\
\textrm{Flex}(S) &= (
  \{f : \mathbb{R}_{\geq 0} \to S \mid f \textrm{ non zeno} \}
  ,
  \cdot
  )
\end{align*}
While the type of rigid variables over a base set is computed
according to the functor
\begin{align*}
\textrm{Rigid} &: \textrm{Set} \to Psh(\text{\sf B}\mathcal{F})\\
\textrm{Rigid}(S) &= (S,((\_,x) \mapsto x))
\end{align*}

There is a natural inclusion morphism from
$\texttt{Rigid} \to \texttt{Flex}$, which is (for every set)
monic. However, $\texttt{Rigid}(S)$ is not the only subobject of
$\texttt{Flex}(S)$. Any stuttering- and translation-closed subset of
behaviors will be interpretable as a temporal set. Of course, these
are not the only temporal types: the inner hom between types of
flexible variables, for instance, corresponds to temporal processes
rather than flexible variables over functions of the underlying sets.

In Section~\ref{sec:hol-background} we reviewed the fact that
a modal hyperdoctrine may be represented by applying the direct image
part of the geometric morphism to the subobject classifier in
$Psh(\text{\sf B}\mathcal{S})$.  As $\mathcal{S}$ is a group, it has
only two ideals, $\emptyset$ and $\mathcal{S}$. Thus,
$\Omega_{Psh(\text{\sf B}\mathcal{S})}$ is the set $\mathbb{2}$ with
the trivial $\mathcal{S}$-action.

As presheaf categories have all (co)limits, the direct image
part of the geometric morphism may be computed as a right Kan
extension. As our categories $\text{\sf B}\mathcal{S}$ and $\text{\sf
B}\mathcal{F}$ have singleton objects,
this can be computed
pointwise. Given $\text{\sf F}:\textbf{\sf Set}^{\text{\sf
B}\mathcal{S}}$, we compute
$\textrm{lim}\left(\bullet_{\mathcal{S}}\downarrow\iota\stackrel{\pi^{\bullet_{\mathcal{S}}}}{\to}\text{\sf
B}\mathcal{S}\stackrel{\text{\sf F}}{\to}\textbf{\sf Set}\right)$,
which amounts to equalizing away the stutter action
\begin{tikzcd}
\prod_{_{\mathcal{S}\setminus}\mathcal{F}}\text{\sf F}(\bullet_{\mathcal{S}})\arrow[tail]{r}&\prod_{\mathcal{F}}\text{\sf F}(\bullet_{\mathcal{S}})\arrow[shift  left=1.5ex]{r}\arrow{r}&\prod_{\mathcal{S}\times\mathcal{F}}\text{\sf F}(\bullet_{\mathcal{S}})
\end{tikzcd}.\\
On $Psh(\text{\sf B}\mathcal{S})$'s subobject classifier, this is
\begin{align*}
\textrm{Prop} &\triangleq \iota_\star (\Omega_{Psh(\text{\sf B}\mathcal{S})}) \\
              &= (\{p : \mathcal{F} \to \mathbb{2} \mid \forall s \in \mathcal{S}, f \in \mathcal{F}, p(f) = p(f\cdot s)\}\\
  &\quad \quad , \quad (f,p) \mapsto (f' \mapsto p(f \cdot f'))) \\
&\cong (\mathcal{P}(\mathbb{R}_{\geq 0}),(f,O) \mapsto \textrm{im}^{-1}(f)(O))
\end{align*}

Consequently (and pleasingly), in our model, a proposition corresponds
to the set of times when that proposition is true.

All the usual connectives coming from the boolean algebra structure
are computed pointwise.  All that remains is to compute the modal
structure.  The subobject classifier in $Psh(\text{\sf B}\mathcal{F})$
is the collection of falter ideals \[\Omega_{Psh(\text{\sf
B}\mathcal{F})}=\{I\subseteq\mathcal{F}\mid\forall i\in I~\forall
f\in\mathcal{F}.~i\cdot f\in I\},\] but these are just all
upward-closed subsets of $\mathbb{R}_{\geq 0}$, so
$\Omega_{Psh(\text{\sf
B}\mathcal{F})}\cong\langle \mathcal{P}_{\uparrow}(\mathbb{R}_{\geq
0}),(n,O) \mapsto \textrm{im}^{-1}(n)(O) \rangle$.  As subobject
classifier in $Psh(\text{\sf B}\mathcal{F})$, $\Omega_{Psh(\text{\sf
B}\mathcal{F})}$ is initial in complete Heyting algebras internal to
$\mathcal{F}$, so the obvious equivariant inclusion
$i_{\Omega}:\Omega_{Psh(\text{\sf
B}\mathcal{F})}\hookrightarrow\iota_{\star}(\Omega_{Psh(\text{\sf
B}\mathcal{S})})$ is essentially unique. The right adjoint
$\tau_{\Omega}:\iota_{\star}(\Omega_{Psh(\text{\sf
B}\mathcal{S})})\to\Omega_{Psh(\text{\sf B}\mathcal{F})}$,
which classifies $1\to\iota_{\star}(\Omega_{Psh(\text{\sf
B}\mathcal{S})})$, is, then, the upward closure
$\uparrow(\--):\mathcal{P}(\mathbb{R})\to\mathcal{P}_{\uparrow}(\mathbb{R})$.
The adjunction
$\Box:=i_{\Omega}\circ\tau_{\Omega}:\textrm{End}(\iota_{\star}\Omega_{Psh(\text{\sf
B}\mathcal{S})})$ provides a left exact comonad on the complete
internal Heyting algebra $\iota_{\star}(\Omega_{Psh(\text{\sf
B}\mathcal{S})})$.

The resulting modal structure is quite natural -- it reduces to
ensuring that a proposition holds at all future times
\begin{align*}
\Box(-) &: \textrm{Prop} \to_{Psh(\text{\sf B}\mathcal{F})} \textrm{Prop}\\
\Box(S) &= \{r \in \mathbb{R}_{\geq 0} \mid \forall r' \geq r, r' \in S\}.
\end{align*}
As such, our categorical model is precisely a higher order generalization of the
continuous-time semantics presented in Section~\ref{sec:tla-background}.
\begin{theorem}
  The modal hyperdoctrine $(Psh(\text{\sf
  B}\mathcal{F}), \textrm{Hom}(\--,\iota_{\star}(\Omega_{\text{\sf
  B}\mathcal{S}})))$ admits a sound
  interpretation of higher-order classical S4.  Moreover, restricting
  to the first-order fragment, this model corresponds to the model of
  the Temporal Logic of Actions in
  Figure~\ref{fig:tla-continuous-semantics} and agrees for validity
  with the standard semantics (Figure~\ref{fig:tla-lamport-semantics}).
\end{theorem}
\section{Conclusion}
\label{sec:conclusion}

We have found a categorical setting in which to model a higher-order
version of TLA, providing a way of assigning meaning to statements in
this logic. This a first step towards a useful higher-order temporal
logic for digital systems. In particular, the model we have described
will allow us to formulate proof rules and verify that
they are sound with respect to our model. We imagine that other
models for such a proof theory may also be of interest.

Our model construction started by switching from the discrete-time
semantics for TLA that was originally formulated by Lamport to a
real-time semantics. This was essential, since stuttering invariance
does not correspond to closure under a group action in the discrete
case. In Lamport's semantics, stuttering forms a monoid (at best)
rather than a group, and closure under the action of that monoid fails
to fully account for stuttering invariance. Nonetheless, a categorical
semantics of higher-order TLA based on discrete-time stuttering
invariance remains an intriguing challenge.

We plan to continue our work on a higher-order TLA, with the goal of
using it as the basis of a proof assistant and toolchain for
practical engineering purposes. Yet significant challenges remain,
such as developing the required syntax, proof theory, and so
on. Moreover, it remains to be seen how extending TLA with
higher-order features can be put into useful practice. A potential use
case would be to specify a variant of
PlusCal \cite{lamport2009pluscal}, a programming language that
translates to TLA, then extending it with handy higher-order features such
as closures or objects.

Our goal in this paper was to find \emph{a} model satisfying our
desiderata. It remains to state what, exactly, ``higher-order TLA'' is
and to specify its class of models. In the present paper we focused on
giving an account of the temporal types, neglecting the underlying
non-temporal sets. A detailed and generalized account of the
categorical properties of TLA's action lifting construction will
necessarily be needed in future work.  All that said, the particular
form of the model we found is intriguing. Because the underlying category
of our hyperdoctrine is a topos, and not just cartesion closed, it has
all finite limits. As such, it is a promising setting for developing
an account of specification composition using
pullbacks \cite{goguen1991categorical,DBLP:conf/fmics/Johnson-FreydHA16}.

\bibliographystyle{eptcs}
\bibliography{act2019}

\onecolumn\newpage
\appendix

\section{Rules of Higher-order logic}
\label{sec:ihol-rules}

\begin{figure}[h]
  \begin{gather*}
    \begin{aligned}
      T,S \in \textrm{Types} &::= \ldots \mid T \to S \mid \textrm{Prop}~~~~
      M,N,O \in \textrm{Terms} &::= \ldots \mid x \mid \lambda (x : T).M
      \mid M~N \mid (\Rightarrow) \mid \forall_T 
   \end{aligned}\\
  \begin{aligned}
    M \Rightarrow N &\triangleq (\Rightarrow)~M~N\\
    \forall (x : T).M &\triangleq \forall_T (\lambda (x : T).M) \\
    \bot &\triangleq \forall (p : \textrm{Prop}).p\\
    \top &\triangleq \forall (p : \textrm{Prop}).p \Rightarrow p
  \end{aligned}\quad \quad
  \begin{aligned}
    \lnot M &\triangleq M \Rightarrow \bot \\
    M \land N &\triangleq \forall (p : \textrm{Prop}).(M \Rightarrow N \Rightarrow p) \Rightarrow p\\
    M \lor N &\triangleq \forall (p : \textrm{Prop}).(M \Rightarrow p) \Rightarrow (N \Rightarrow p) \Rightarrow p\\
    \exists (x : T).M &\triangleq \forall (p : \textrm{Prop}).(\forall (x:T).M \Rightarrow p) \Rightarrow p
  \end{aligned}
\end{gather*}
\begin{mathpar}
\inferrule{\Gamma \vdash M \equiv N : T}{\Gamma \vdash N \equiv M : T}~~~~
\inferrule{
        \Gamma \vdash M \equiv N \\
        \Gamma \vdash N \equiv O
        }{
        \Gamma \vdash M \equiv O
        }~~~~
\inferrule{(x : T) \in \Gamma}{\Gamma \vdash x \equiv x : T}~~~
~~~~\\
\inferrule{
        \Gamma \vdash M_1 \equiv M_2 : S \to T\\
        \Gamma \vdash N_1 \equiv N_2 : S
        }{
        \Gamma \vdash M_1~N_1 \equiv M_2~N_2 : T
        }~~~~
\inferrule{
\Gamma, x : T \vdash M \equiv N : S
}{
\Gamma \vdash \lambda (x : T).M \equiv \lambda (x : T).N : T \to S
}\\
\inferrule{\Gamma, x : T \vdash M \equiv M : S \\
\Gamma \vdash N \equiv N : T
}{
\Gamma \vdash (\lambda (x : T).M)~N \equiv M\msubst{N/x} : S
}~~~~
\inferrule{
 \Gamma, x : T \vdash M~x \equiv N~x : S
}{
 \Gamma \vdash M \equiv N : T \to S
}~~~
\\
\inferrule{ }{\Gamma \vdash (\Rightarrow) \equiv (\Rightarrow) :
  \textrm{Prop}\to\textrm{Prop}\to\textrm{Prop}}~~~
\inferrule{ }{\Gamma \vdash \forall_T \equiv \forall_T : (T \to \textrm{Prop}) \to \textrm{Prop}} \\
\inferrule{ }{\Gamma \mid \emptyset \vdash \textbf{wf}}~~~~
\inferrule{
\Gamma \mid \Theta \vdash \textbf{wf}\\
\Gamma \vdash M \equiv M : \textrm{Prop}
}{
\Gamma \mid \Theta, M \vdash \textbf{wf}
}~~~
\inferrule{
M \in \Theta \\
\Gamma \mid \Theta \vdash \textbf{wf}
}{
\Gamma \mid \Theta \vdash M~\textbf{true}
}\\
\inferrule{
\Gamma \mid \Theta \vdash M~\textbf{true}\\
\Gamma \vdash M \equiv N : \textrm{Prop}
}{
\Gamma \mid \Theta \vdash N~\textbf{true}
}~~~~
\inferrule{
\Gamma \mid \Theta \vdash M \Rightarrow N~\textbf{true}\\
\Gamma \mid \Theta \vdash M~\textbf{true}
}{
\Gamma \mid \Theta \vdash N~\textbf{true}
}~~~~
\inferrule{
\Gamma \mid \Theta, M \vdash N~\textbf{true}
}{
\Gamma \mid \Theta \vdash M \Rightarrow N
}\\
\inferrule{
\Gamma \mid \Theta \vdash \forall_T~M~\textbf{true}\\
\Gamma \vdash N \equiv N : T
}{
\Gamma \mid \Theta \vdash M~N~\textbf{true}
}~~~~
\inferrule{
\Gamma \vdash M \equiv M : T \to \textrm{Prop}\\
\Gamma \mid \Theta \vdash \textbf{wf} \\
\Gamma, x : T \mid \Theta \vdash M~x~\textbf{true}
}{
\Gamma \mid \Theta \vdash \forall_T~M~\textbf{true}
}
\end{mathpar}
\caption{Intuitionistic Higher-order Logic}
\label{fig:iHOL}
\end{figure}

\end{document}